\documentclass[letter]{aa} 
\usepackage{graphicx}
\usepackage{txfonts}
\usepackage{natbib}
\bibpunct{(}{)}{;}{a}{}{,}
\newcommand{\solm}{M$_{\odot}$}

\begin{document}
   \title{The possibility of detecting Sagittarius~A* at $\rm
   8.6\,\mu$m}

   \subtitle{Sensitive, high-resolution imaging observations of the
   Galactic Center}

   \author{R. Sch{\"o}del
          \inst{1}
          \and
          A. Eckart
          \inst{1,2}
          \and
          K. Mu\v{z}i\'{c}
          \inst{1}
          \and
           L. Meyer
          \inst{1}
          \and
          T. Viehmann
          \inst{1}
          \and
          G. C. Bower
          \inst{3}
          }

   \offprints{R. Sch{\"o}del}

   \institute{I.Physikalisches Institut, Universit\"at zu K\"oln, 
              Z\"{u}lpicher Str.1, D-50937 K\"oln\\
              \email{rainer,eckart,muzic,meyer@ph1.uni-koeln.de}
         \and
          Max-Planck-Institut f\"ur Radioastronomie, Auf dem
              H\"ugel 69, D-53121 Bonn
         \and
         Astronomy Department and Radio Astronomy Laboratory,
              University of California, Berkeley. CA 94720
              }

   \date{Received ; accepted }

 
  \abstract 
{Sagittarius~A* (Sgr~A*) at the center of the Milky Way is a black
   hole accreting at extremely sub-Eddington rates. Measurements of its
   emission in the infrared and X-ray domains are difficult due to its
   faintness and high variability.}
{The Galactic Center was observed at 8.6\,$\mu$m in order to detect a
  mid-infrared (MIR) counterpart of Sgr~A*, parallel to NIR
  observations. The goal was to set constraints on possible emission
  mechanisms.}
{Imaging data were acquired with the adaptive optics assisted
  NIR instrument NACO and the MIR instrument VISIR
  at the ESO VLT.}  
{We present MIR imaging data of an unprecedented quality in terms of
  spatial resolution and sensitivity. An extended ridge of emission is
  found to be present in the immediate vicinity of Sgr~A* and thus
  renders any detection of a point source difficult. No MIR point
  source related to Sgr~A* was detected during the observations. We
  derive a tight upper limit of $22\pm14$\,mJy (dereddened) on any
  possible point source present during the observations in the night
  of 4/5 June 2006. The absence of a flare in simultaneous
  observations at 2.2\,$\mu$m and the low limits on any possible
  variability in the MIR suggest strongly that Sgr~A* was in a
  quasi-quiescent state during this night. During the night from 5 to
  6 June 2006, Sgr~A* was found to be variable on a low level at
  3.8\,$\mu$m. No point source at 8.6\,$\mu$m was detected during the
  simultaneous MIR observations. Due to the poorer atmospheric
  conditions a higher upper limit of 60$\pm$30\,mJy was found for
  Sgr~A* at 8.6\,$\mu$m during the second night. }
{ The observations are consistent with theoretical predictions. If the
  published models are correct, the observations demonstrate
  successfully that a 8.6\,$\mu$m counterpart of Sgr~A* can be easily
  detected in its flaring state. Spectral indices derived from
  simultaneous observations of flaring emission from Sgr~A* at NIR and
  MIR wavelengths will enable us to distinguish between different
  kinds of flare models.}


   \maketitle
%

\section{Introduction}

The center of the Milky Way harbors a supermassive black hole of
$3.6\times10^{6}$\solm~\citep[e.g.,][]{Schoedel2002Natur,Ghez2003ApJ,Eisenhauer2005ApJ}.
The non-thermal source related to this supermassive black hole,
Sagittarius~A* (Sgr~A*), radiates at only $10^{-9}-10^{-10}$ times its
Eddington luminosity from radio wavelengths to the X-ray domain. Its
low luminosity is consistent with emission from so-called radiatively
inefficient accretion flows, a jet or a combination of the two models
\citep[e.g.,][]{Yuan2002A&A,Yuan2003ApJ,Bower2004Sci,Shen2005Natur}.

X-ray and near-infrared (NIR) counterparts to Sgr~A* were only
discovered with the availability of sensitive, high-resolution
instruments for these wavelengths. It was found that Sgr~A* is highly
variable at these wavelengths, showing flaring emission on time scales
of $\sim60-100$\,min, with flux increases up to 100 at X-rays and up
to 10 in the NIR
\citep[e.g.,][]{Baganoff2001Natur,Genzel2003Natur}. The variability at
X-ray and IR wavelengths appears to be simultaneous
\citep{Eckart2004A&A,Eckart2006A&A}. At MIR wavelengths, only upper
limits to the flux of Sgr~A* have been reported so far
\citep{Stolovy1996ApJ,Telesco1996ApJ,Cotera1999ASPC,Eckart2006A&A}.
The detection of Sgr~A* at MIR wavelengths is difficult due to the
lower spatial resolution compared to NIR wavelengths, the general
difficulties of imaging in the thermal IR regime, and the presence of
warm dust near Sgr~A*. Warm dust is associated with the mini-spiral
gas streamers that pass close to Sgr~A*. Therefore, Sgr~A* is no
isolated point source in the MIR and its detection requires high image
quality, above all high spatial resolution, in order to achieve a
sufficiently high contrast. Here, we report on new MIR observations,
using the European Southern Observatory's MIR imager and spectrograph
VISIR at the Very Large Telescope (VLT) on Cerro Paranal in Chile.

Although Sgr~A* was not detected, the acquired images are -- in terms
of sensitivity and spatial resolution -- the highest quality
$8.6$\,$\mu$m-images of the Galactic Center (GC) region published up
to now, allowing us to report the so far tightest upper limit on the
$8.6$\,$\mu$m flux of Sgr~A*. Even more interesting is that we can
show -- via simultaneously acquired adaptive optics NIR imaging data
-- that the infrared flux of Sgr~A* agrees well with models of the
quiescent/low activity emission.  This information allows us to
conclude that --according to currently accepted theoretical models --
Sgr~A* can be easily detected at MIR wavelengths during a bright
flare. Such an observation will allow to derive the NIR-to-MIR
spectral slope of Sgr~A* during flares and thus to distinguish between
different flare models.


\section{Observations and calibration}

   \begin{figure}[!t]
   \centering
   \includegraphics[width=0.72\columnwidth]{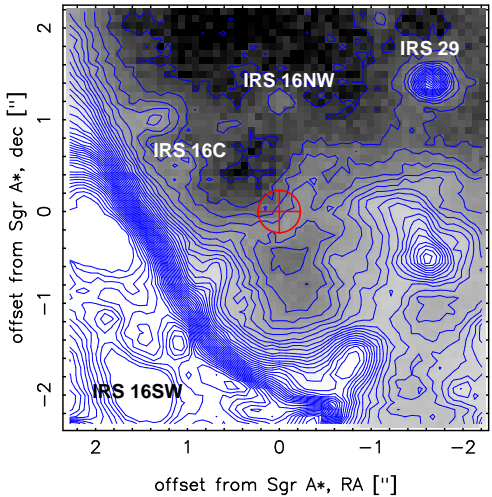}
   \caption{\label{Fig:Obs} Sgr~A* and its immediate environment at
   $8.6$\,$\mu$m.  Direct shift-and-add image.  Contour lines as in
   are plotted in steps of 0.5\,mJy from 0.5 to 20\,mJy per pixel (one
   pixel corresponds to $0.075''\times0.075''$).  Some well known
   sources are labeled.  The red circle has a radius corresponding to
   the FWHM of the diffraction limited beam size ($\sim$$0.25''$) and
   is centered on the position of Sgr~A*. The astrometric accuracy is
   $\leq0.03''$.}
   \end{figure}

The GC was observed in the MIR N-band with the ESO VLT unit telescope
3 (UT~3) using the MIR imager/spectrograph VISIR
\citep{Lagage2003SPIE,Lagage2004Msngr} in the nights of 4/5 and 5/6
June 2006. The pixel scale was $0.075''$ per pixel. The PAH\,1 filter
with a central wavelength of $8.59$\,$\mu$m and half-band width of
$0.42$\,$\mu$m was used. The standard nodding ($\sim$$330\deg$ east of
north) and chopping (with a chop throw of $30''$) technique was used.
Standard data reduction was applied, i.e.\ the sky background acquired
during the chopping and nodding observations was subtracted from the
images of the target and individual frames were combined with a simple
shift-and-add technique.   Dithering was applied during the
observations, leading to a FOV in the combined mosaic images of
$\sim$$30''\times30''$. The seeing during the observations on 5 June
2006 was good enough to result in a PSF FWHM of $\sim$$0.3''$, close
to the diffraction limit of the VLT at
8.6\,$\mu$m.

In this work, we use also NIR imaging data that were obtained parallel
to the MIR data with the adaptive optics imager/spectrograph NACO at
the ESO VLT. On June 5, the data were obtained in the Ks-band in
polarimetric mode with a pixel scale of $0.013''$ per pixel. The data
reduction and flux calibration was identical to the procedures
described in \citet{Eckart2006A&Aa} and in Meyer et al.\ (2006, A\&A,
in press). The L'-band ($3.8$\,$\mu$m) imaging data were obtained with
NACO/VLT on 29~May and 6~June 2006 with a pixel scale of $0.027''$ per
pixel. The reduction of the L'-band data was standard and identical to
the data reduction procedures described in Muzic et al.\ (2006,
submitted to A\&A). Details of the NIR  and MIR observations are
listed in Tab.~\ref{Tab:Obs}.

   \begin{table*}
   \centering 
         \begin{tabular}{llllllllll}
            \hline
            \noalign{\smallskip}
            Date   &  $\rm t_{start}$ [UTC] &  $\rm t_{stop}$ [UTC] & $\lambda_{c}$ [$\mu$m] & $\Delta\lambda$ [$\mu$m] &
            $DIT$ [s] & $NDIT$ & $N$ & seeing [$''$] \\
            \noalign{\smallskip}
            \hline
            \noalign{\smallskip}
            29 May 2006 &  04:55 &  08:01 & 3.80 & 0.62 & 0.2 & 150 & 108 & $0.6''$ \\
            05 June 2006 &  04:55 &  08:01 & 8.59 & 0.42 & 0.02 & 98 & 1104 & $0.6''$ \\
            06 June 2006 &  04:33 &  07:37 & 8.59 & 0.42 & 0.02 & 98 & 1104 & $1.5''$ \\
            05 June 2006 &  04:55 &  08:01 & 2.18 & 0.35 & 30 & 1 & 62 & $0.6''$ \\
            06 June 2006 &  04:33 &  07:37 &  3.80 & 0.62 & 0.2 & 150 & 1104 & $1.5''$ \\
            \noalign{\smallskip}
            \hline
         \end{tabular}
     \caption{\label{Tab:Obs} Summary of observations with VISIR and
      NACO at the ESO VLT used in this work. $\lambda_{c}$ is the
      central wavelength of the filter used and $\Delta\lambda$ its
      FWHM. DIT is the detector integration time, NDIT the number of
      detector readouts averaged online, and N the total number of
      images of the target that were acquired. The total integration
      time corresponds to $DIT\times NDIT\times N$. The total
      integration time during both MIR observations was thus
      36\,min. The last column indicates the average optical seeing
      conditions during the observations.}
   \end{table*}

Flux calibration was achieved by observations of the standard star
HD178345 \citep[$14.32$\,Jy in the PAH\,1 filter, see ESO VISIR web site
and ][]{Cohen1999AJ}. The isolated standard star was also used as PSF
reference for image deconvolution.

   \begin{figure}[!b]
   \centering
   \includegraphics[width=0.72\columnwidth]{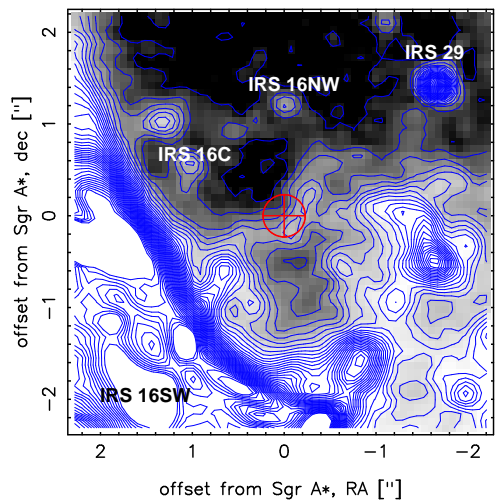}
   \caption{\label{Fig:Deconv} Lucy-Richardson deconvolved and beam
   restored version of the image shown in Fig.~\ref{Fig:Obs}, with
   identical labeling and contour lines. We derive fluxes of
   $22\pm5$, $21\pm5$, and $180\pm20$~mJy for the point sources
   IRS~16C, IRS~16NW, and IRS~29 (without extinction correction).  The
   flux of Sgr~A* was estimated by subtracting sources of given flux
   at the position of Sgr~A* from the image in Fig.~\ref{Fig:Obs} and
   checking whether a deformation of the dust ridge at the position of
   Sgr~A* could be seen in the deconvolved and beam restored image.}
   \end{figure}

The astrometric reference frame could be established via the positions
and proper motions of the stars IRS~3, IRS~7, IRS~9, IRS~14NE,
IRS~12N, IRS~2L, IRS~6E, and IRS~29, that were published by
\citet{Ott2004PhDT}. All of these stars were detected on the VISIR
image as point sources. The positions of these eight stars were used
to solve transformation equations up to second order. The accuracy of
the astrometric solution could be checked by choosing sub-groups of
seven out of the eight stars and repeating the transformation. Also,
we compared the measured astrometric position of the stars IRS~16NW,
IRS~16C, and IRS~29 to their predicted positions.  The position of
Sgr~A* could thus be established with a $1 \sigma$ uncertainty of
$0.03''$ (less than half a pixel) in the MIR data. This is about a
factor of 10 better than in previous work
\citep{Stolovy1996ApJ}. There are two main reasons for this increased
astrometric accuracy: on the one hand, the high sensitivity and
spatial resolution of the VISIR data, and on the other hand, the
improved IR position of Sgr~A* due to precise positions and proper
motions of stars in the NIR
\citep[e.g.,][]{Genzel2000MNRAS,Ott2004PhDT} and due to the well-known
position of Sgr~A* in the IR reference frame via SiO maser sources
\citep[e.g.,][]{Reid2003ApJ}.

{\bf Figure~\ref{Fig:Obs} presents the flux calibrated direct
shift-and-add image image of the central arcseconds around Sgr~A* for
the night of 4/5 June 2006.  Significant signal power was found to be
present at the diffraction limit ($>60\%$, estimated by aperture
photometry).}

\section{Flux limits on Sgr~A* at $\mathbf{8.6}$, $\mathbf{3.8}$, and
  $\mathbf{2.2}$ \,$\rm \bf \mu$m}

   \begin{figure}[!b]
   \centering
   \includegraphics[angle=270,width=0.72\columnwidth]{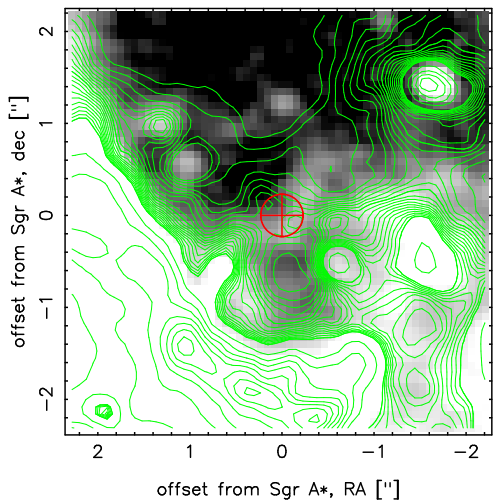}
   \caption{ \label{Fig:NL} Contours of diffuse L'-band emission
   overplotted onto the VISIR $8.6$\,$\mu$m image. Contour levels are
   drawn between 10 and 200\,$\mu$Jy in steps of 10\,$\mu$Jy and
   between 0.1 and 2\,mJy in steps of 0.1\,mJy}
   \end{figure}

A Lucy-Richardson deconvolved and beam restored version of the image
from Fig.~\ref{Fig:Obs} is shown in Fig.~\ref{Fig:Deconv}. The
deconvolved and beam restored image resembles closely the original
image, but emphasizes some of the finer details. There is a dust ridge
present very close to the position of Sgr~A*, where it bends from an
east-west extension to a direction southeast-northwest. In earlier MIR
observations of the GC \citet{Stolovy1996ApJ} have found very similar
structures near Sgr~A*. Due to their high spatial resolution and
sensitivity, the new data show much richer details, however. Even some
stars, such as IRS~29, IRS~16NW and IRS~16C (labeled in
Fig.~\ref{Fig:Obs}) can be detected.  In spite of the high quality of
the data, no obvious point source is present at the position of
Sgr~A*.  We conclude that the source coincident with Sgr~A* that was
reported by \citet{Stolovy1996ApJ} was most probably related to the
background emission of the dust ridge and not to Sgr~A* itself.  All
the flux appears to originate from the dust ridge. This assumption can
be used to obtain a first simple flux estimate for Sgr~A*. Such an
estimate can be obtained by adding or subtracting point sources of
given flux at the position of Sgr~A* and checking whether the shape of
the dust ridge is altered perceptibly in the images. After testing
various flux levels, we estimate that a point source at the position
of Sgr~A* that is present in the image cannot be brighter than
$\sim$5\,mJy.

   \begin{figure}[t]
   \centering \includegraphics[width=\columnwidth]{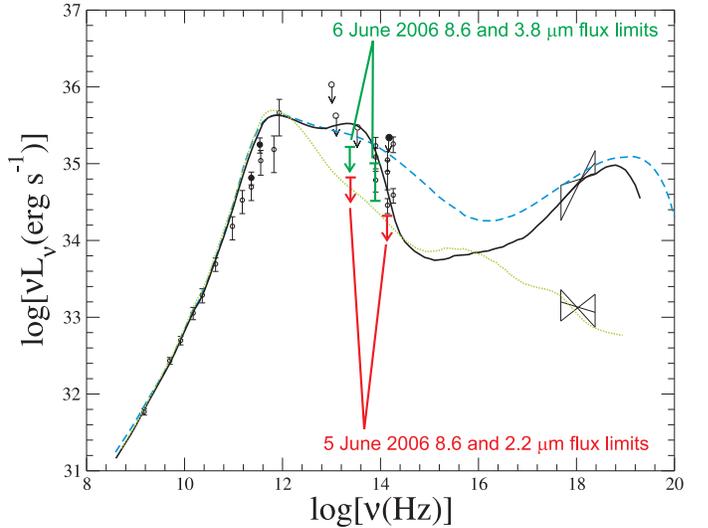}  
   \caption{\label{Fig:Models} Emission models for Sgr~A*
 \citep{Yuan2003ApJ,Yuan2004ApJ}.  The upper limit on the flux of
 Sgr~A* at $8.6$\,$\mu$m determined in this work is indicated together
 with the simultaneously measured (NACO/VLT) flux at $3.8$\,$\mu$m on
 June 6 and the upper limit on the 2.2\,$\mu$m emission on June 5
 superposed onto RIAF models of the quiescent (dotted line) and
 flaring (dashed and solid lines) emission from Sgr A*. All flux
 measurements were corrected for extinction. The spectral slope
 measured between the MIR and NIR during a flare will allow to
 distinguish between a model where the X-ray emission is due to
 acceleration of electrons that produce NIR synchrotron radiation that
 is up-scattered into the X-ray domain (steep MIR-to-NIR slope), and a
 model where the NIR emission is due to synchrotron emission from
 electrons heated to a higher temperature, while the X-ray emission is
 due to synchrotron emission from some electrons that were accelerated
 into a hard power law distribution (flat MIR-to-NIR slope). }
   \end{figure}

With AO L'-band imaging \citet{Eckart2006A&A} and \citet{Ghez2005ApJ}
have demonstrated the presence of a dust blob between about $0.03''$
and $0.15''$ southwest of Sgr~A*. As discussed by
\citet{Eckart2006A&A}, it is probably this blob that is responsible
for most of the MIR emission near the position of Sgr~A* when the
latter is in a quiescent state. This blob is clearly visible in the
MIR images presented here and forms part of the dust ridge.
Therefore, a second way of estimating the flux of a putative point
source near Sgr~A* can be obtained by trying to account for the
extended dust emission. For this purpose we used a high-quality
L'-band image obtained with NACO in the night 29/30 May 2006. The
\emph{StarFinder} \citep{Diolaiti2000A&AS} code was used to extract
the stars from the image (including a point source due to the emission
from Sgr~A*) and thus to obtain an image of the diffuse emission at
$3.8$\,$\mu$m. This image was smoothed, transformed and scaled (with
the help of the IDL routines POLYWARP and POLY\_2D) to fit the scale,
orientation, and resolution of the MIR data.  Flux calibration of the
L'-band data was achieved by using the published fluxes of the sources
IRS~16C and IRS~33N \citep{Blum1996ApJ}. Figure~\ref{Fig:NL} shows the
diffuse $3.8$\,$\mu$m flux superposed as contour lines onto the MIR
image from the right panel of Fig.~\ref{Fig:Obs}. The median ratio of
the diffuse flux at the two wavelengths in a circular area of $0.75''$
around Sgr~A* is $29\pm7$. This value was used to scale the diffuse
flux present in the L'-band data and to subtract it from the MIR
image. Here it is important to note that the flux density ratio is
much higher in other areas of the mini-spiral. In the northern arm it
can be up to 100. Therefore we obtain a conservative estimate of the
remnant flux. After subtraction of the scaled diffuse $3.8$\,$\mu$m
emission we find an upper limit on the flux of a putative point source
at the position of Sgr~A* of 8$\pm5$\,mJy. Following
\citet{Lutz1996A&A}, this corresponds to an extinction corrected flux
of $22\pm14$\,mJy for the image obtained on 5 June 2006. The lower
data quality on 6 June 2006 leads to a higher upper limit of
$60\pm30$\,\,mJy (extinction corrected) on any possible Sgr~A*
counterpart.

During the simultaneous observations at $2.2$\,$\mu$m on June 5, no
counterpart to Sgr~A* was detected. The upper limit for the flux of
Sgr~A* was determined to $2\pm1$\,mJy. Although the quality of the
L'-band observations from June 6 was fairly low (due to bad seeing,
poor AO correction and electronic noise on the detector), a weak
variable counterpart of Sgr~A* could be detected. Its variability was
measured to be a factor $\lesssim 2$ and its extinction
\citep{Lutz1996A&A} corrected flux to be $10\pm5$\,mJy.

\section{Discussion}

A central question is whether Sgr~A* was really in a state of
quasi-quiescence when the MIR observations were taken or whether
Sgr~A* is generally too faint to be detected at MIR wavelengths, even
when a flare occurs. We argue that Sgr~A* was in a state of
quasi-quiescence during the observations  on 5 June 2006 for
three reasons. (a) The simultaneous observations with NACO at
$2.2$\,$\mu$m show that Sgr~A* was not detected at $2.2$\,$\mu$m,
i.e.\  no NIR flare was observed. (b) Models predict the emission of
Sgr~A* at $8.6$\,$\mu$m during a flare to be up to 10 times stronger
than the upper limit reported by us (see Fig.~\ref{Fig:Models}). (c)
Differential imaging allows us to derive tight upper limits on the
variability of Sgr~A* at $8.6$\,$\mu$m during the observations. From a
difference image between the data from 5 June and 6 June an upper
limit of $15\pm10$\,mJy can be derived on any possible variability of
Sgr~A* between these two days. The high quality data from 5 June allow
to estimate the upper limit to the variability of Sgr~A* during the
observations from differential imaging between individual images
($\sim$30\,s of integration time, with 1.5 to 2.5\,min time between
them) taken during this night to $10\pm6$\,mJy.   Our conclusion
is that Sgr~A* was indeed in a state of quasi-quiescence during the
MIR observations at $8.6$\,$\mu$m.

The lower quality of the June 6 data leads to higher limits of
$30\pm20$\,mJy for variability between the individual images. The
L'-band observations show that Sgr~A* was continuously variable during
the observations on 6 June. The simultaneously acquired upper limit on
the average flux of Sgr~A* at 8.6\,$\mu$m and the limits on shorter
timescale variability at this wavelength are consistent with the
measurements at 3.8\,$\mu$m

Measurements of the Sgr~A* quiescent and flaring emission are
indicated in Fig.~\ref{Fig:Models} along with theoretical models
\citep[data and models taken from ][]{Yuan2003ApJ,Yuan2004ApJ}. The
simultaneous measurements at $2.2$, $3.8$, and $8.6$\,$\mu$m derived
in this work are indicated in the Figure. Two conclusions can be
drawn. (a) The acquired upper limits on the IR flux density of Sgr~A*
at $2.2$ and $8.6$\,$\mu$m set tight constraints on the
quasi-quiescent emission and are consistent with the theoretical
models shown in the figure.  The measurements at $3.8$ and
$8.6$\,$\mu$m are consistent with the models, too. The measurements
are also in agreement with the synchrotron/SSC models presented by
\citet{Eckart2006A&A}. Published jet models predict significantly
higher fluxes in the MIR
\citep[see][]{FalckeMarkoff2000A&A,Yuan2002A&A}, but can probably be
easily adapted to account for the new upper limits. (b) VISIR at the
ESO VLT is sensitive enough to detect flaring emission from Sgr~A*
during a bright flare without difficulty and during a fainter flare
during periods of good and stable atmospheric conditions. When
combined with simultaneously acquired NIR measurements the MIR-to-NIR
spectral index derived from such observations can be used to
distinguish between different flare models.

\begin{acknowledgements}
      Part of this work was supported by the German \emph{Deut\-sche
      For\-schungs\-ge\-mein\-schaft DFG\/} Sonderforschungsbereich
      project number SFB 494. We thank the referee for her/his helpful
      comments.
\end{acknowledgements}

\bibliographystyle{aa}

\bibliography{gc}

\end{document}